\documentclass[mreferee]{gji}
\usepackage{timet}

\title[Friction laws from dimensional-analysis point of view]
{Friction laws from dimensional-analysis point of view}
\author[T. Hatano]
  {Takahiro Hatano$^1$\\
  $^1$ Earthquake Research Institute, The University of Tokyo,
    Yayoi, Bunkyo \emph{1130032} Tokyo, Japan}
\date{}
\pagerange{\pageref{firstpage}--\pageref{lastpage}}
\volume{}
\pubyear{}

\begin{document}
\label{firstpage}
\maketitle

\begin{summary}
Friction laws, which are a key to the understanding of the diversity of earthquakes, are considered theoretically. 
Using dimensional analysis, the logarithmic dependence of the friction coefficient on the slip velocity and the state variable is derived without any knowledge of the underlying physical processes on the frictional surface.
This is based on a simple assumption that the friction coefficient is expressed as the difference from a reference state.
Therefore, the functional form of the rate and state dependent friction law itself does not necessarily mean that thermal activation processes dominate friction.
It is also shown that, if there are two (or more) state variables having the same dimension, we need not assume the logarithmic dependence on the state variables.
\end{summary}


\section{Introduction}
Numerous attempts have been made to clarify the frictional properties of rocks to understand the diverse nature of the rupture dynamics of earthquakes \cite{kanamori1994}.
They are investigated mainly using laboratory experiments, and some of them are summarized as an empirical law, which is now known as the rate and state dependent friction (RSF) law \cite{dieterich1978,ruina1983}.
A general form of the RSF law describes the friction coefficient $\mu$ as a function of the slip velocity $V$ and the state variables $\{\theta\}=\theta_1,\theta_2,\cdots$.
\begin{equation}
\label{generalRSF}
\mu=\mu(V, \{\theta\}).
\end{equation}
The state variables represent any relevant microscopic features of the slipping interface.
Conventionally, the state variables are assumed to have the dimension of time.
We thus assume that, throughout this paper, the state variables have the dimension of time unless otherwise indicated.

Importantly, the RSF law involves the difference in friction coefficient between two states, and does not state anything about the absolute value of the friction coefficient.
This is apparent in view of a concrete form of the RSF law (with a single state variable).
\begin{equation}
\label{RSF_1state}
\mu(V,\theta)-\mu(V_*,\theta_*)=a\log\frac{V}{V_*}+b\log\frac{\theta}{\theta_*},
\end{equation}
where $V_*$ and $\theta_*$ are for a reference state, and $a$ and $b$ are nondimensional positive constants.

Because the state variables $\{\theta\}$ describe the microscopic properties on the slipping surface, they cannot be regarded as controllable parameters but time-dependent quantities.
Thus, the RSF law in the form of Eq. (\ref{RSF_1state}) must be supplemented with time evolution laws for state variables.
The time evolution of state variables depend on the microscopic physical processes on the frictional surface, and therefore the evolution laws involve the microscopic details of the system. In this paper, the time evolution laws are not discussed.

In most cases, a single state variable version, Eq. (\ref{RSF_1state}), has been successful in fitting experimental data on rock friction if the time evolution law for $\theta$ is chosen suitably \cite{marone1998}.
If an empirical law reproduces experiments well, it often gives a clue to the underlying (microscopic) physics.
Actually, Chester \shortcite{chester1994} and Heslot et al. \shortcite{heslot1994} suggest that the RSF law is a consequence of thermal activation processes at asperities, and some experimental work followed to validate the proposition \cite{nakatani2001}.
In this respect, the logarithmic term with respect to the slip velocity in Eq. (\ref{RSF_1state}) appears natural.

Here we show an alternative view on the RSF law; Eq. (\ref{RSF_1state}) is derived using dimensional analysis.
This means that the logarithmic term itself does not imply any microscopic physical processes.
(Note that this does not mean that thermal activation processes is irrelevant to the RSF law.)
In a similar manner, we also show the generally permissible form of the RSF law in the existence of two or more state variables.

It should be stressed that the present attempt is not only of purely theoretical interest, but may contribute to the better understanding of RSF as well as to the formulation of new friction laws.
The RSF law with multiple state variables has been used to explain some experimental data under hydrothermal conditions \cite{gu1984,blanpied1986,noda2012}, but there is no a priori reason for assuming an equation like (\ref{RSF_1state}) with multiple state variables.
A clear demonstration of this result is given.
In addition, recent numerous laboratory data show that the conventional RSF law should be replaced by a new law at seismic slip velocities (mm/s to m/s) \cite{goldsby2002,han2007,han2011,rice2006}.
We discuss a possible application of the RSF formulation to this regime.

\section{One state-variable case}
Throughout this paper, we wish to focus on empirical friction laws.
Hence, the friction coefficient must be expressed in terms of experimentally accessible quantities only.
This assumption may be natural if one cannot access sufficient microscopic information for physical modeling.

If the friction coefficient depends on the sliding velocity $V$ and the state variable $\theta$ that has the dimension of time, it can be formally written as
\begin{equation}
\label{muVtheta}
\mu=\mu(V,\theta).
\end{equation}
Here $\mu(V,\theta)$ means the friction coefficient that is experimentally measured at a given velocity $V$ and a certain value of the state variable $\theta$.
Although the state variable $\theta$ itself is not directly measurable, one can instead compute $\theta$ from other observable quantities using an evolution law.

\subsection{Some consequences of the $\Pi$ theorem}
Note that the friction coefficient is a nondimensional quantity.
According to the $\Pi$ theorem for dimensional analysis \cite{barenblatt},
any nondimensional quantity must be a function of nondimensional parameters only.
In view of Eq. (\ref{muVtheta}), however, we cannot produce any nondimensional quantities using velocity and time (i.e., the state variable).
Therefore, the right hand side of Eq. (\ref{muVtheta}) cannot be nondimensional unless other relevant parameters come into play.

For example, one can introduce a length constant $L$ that is defined as the characteristic relaxation length for friction.
\begin{equation}
\label{muVthetaL}
\mu = \mu(V, \theta, L).
\end{equation}
Then $V\theta/L$ is the only nondimensional quantity that can be constructed from $V$, $\theta$, and $L$.
From the $\Pi$-theorem, one can write
\begin{equation}
\label{onevariable}
\mu= f \left(\frac{V\theta}{L}\right),
\end{equation}
where $f(\cdot)$ is a scaling function.
Although Eq. (\ref{onevariable}) is consistent with dimensional analysis, it cannot represent the slip velocity dependence of the steady-state friction coefficient.
To see this, note that the state variable at steady state must be proportional to $L/V$, as there are no other constants that have the dimension of time.
Then, obviously, $V\theta/L$ is independent of the sliding velocity.

To express both the steady-state slip velocity dependence and transient behavior, the two arguments in Eq. (\ref{muVtheta}), $V$ and $\theta$, must be nondimensionalized by other dimensional parameters.
Such parameters are usually material constants or other intrinsic parameters of a system.
For example, the slip velocity $V$ and the state variable $\theta$ may be normalized using the sound velocity $V_S$ and the Debye frequency $\omega_D$, respectively.
Then we can write instead of Eq. (\ref{muVtheta})
\begin{equation}
\nonumber
\mu=f \left(\frac{V}{V_S}, \omega_D\theta\right).
\end{equation}
If this nondimensionalisation is appropriate, experimental data for many materials having different sound velocities and Debye frequencies should collapse to a single master curve.
To the best of our knowledge, however, there have been no attempts to demonstrate such a dependence to this date.
To choose an appropriate nondimensionalisation, we must know the essential physical processes behind a phenomenon in focus and, in general, must consider some models based on microscopic level physics.
Such an approach would lead to an alternative form of the friction law without invoking a reference velocity, which is conventionally denoted by $V_*$.
Although there are some promising attempts \cite{chester1994,heslot1994,bar-sinai2014}, this is beyond the scope of the present work.
More detailed discussion on the microscopic nature of friction law will be given elsewhere.

\subsection{Nondimensionalisation with reference state}
\subsubsection{General form}
If one cannot know intrinsic parameters that have the dimensions of velocity and time,
nondimensionalisation of $V$ and $\theta$ can be achieved by introducing a reference state.
Namely, one may use in place of intrinsic constants {\it the slip velocity and the state variable at a reference state}, $V_*$ and $\theta_*$.
Then we can utilise two nondimensional parameters: $V/V_*$ and $\theta/\theta_*$.
The choice of a reference state $(V_*,\theta_*)$ may be arbitrary, but it is convenient to choose $\theta_*$ as the steady-state value at sliding velocity of $V_*$.
Then $\mu(V_*,\theta_*)$ is just the friction coefficient measured at the steady state of sliding velocity $V_*$.
With the aid of a reference state, one can describe the friction coefficient in the form that is compatible with dimensional analysis.

From the physical point of view, the friction coefficient at $(V, \theta)$ must not depend on $(V_*, \theta_*)$.
One thus cannot write $\mu=\mu(V,\theta,V_*,\theta_*)$, but must distinguish $\mu(V,\theta)$ and $\mu(V_*,\theta_*)$.
Then a friction law should be written as
\begin{equation}
\label{generalform}
F(\mu(V, \theta), \mu(V_*,\theta_*)) = f\left( \frac{V}{V_*}, \frac{\theta}{\theta_*}\right),
\end{equation}
where the functional form of $F(\cdot, \cdot)$ is arbitrary.
This is a general form of friction law that does not contain any intrinsic dimensional parameters.
The choice of $F(\cdot, \cdot)$ defines a concrete form of friction law.

\subsubsection{Derivation of logarithmic dependence}
One may describe the friction coefficient as the change from a reference state value;
i.e., $F(x, y)=x-y$.
Then Eq. (\ref{generalform}) leads to
\begin{equation}
\label{start1}
\mu(V,\theta)-\mu(V_*,\theta_*) = f\left(\frac{V}{V_*},\frac{\theta}{\theta_*}\right).
\end{equation}

It can be shown that Eq. (\ref{start1}) is equivalent to the RSF law, Eq. (\ref{RSF_1state}).
Differentiating the both sides of Eq. (\ref{start1}) with respect to $V$, one obtains
\begin{equation}
\label{pdiff_1}
\frac{\partial\mu(V,\theta)}{\partial V}=\frac{1}{V_*}
\frac{\partial}{\partial (V/V_*)} f\left(\frac{V}{V_*}, \frac{\theta}{\theta_*}\right).
\end{equation}
As $V$, $V_*$, $\theta$, and $\theta_*$ can be chosen arbitrarily, one may set $V_*=V$ and $\theta=\theta_*$.
Then Eq. (\ref{pdiff_1}) becomes
\begin{eqnarray}
\label{partialV}
\frac{\partial\mu(V,\theta_*)}{\partial V}=
\frac{1}{V}\left.\frac{\partial f(s_1,s_2)}{\partial s_1}\right|_{\bf s_0}
=\frac{a}{V},\\
a \equiv \left.\frac{\partial f(s_1,s_2)}{\partial s_1}\right|_{\bf s_0},
\end{eqnarray}
where the scaling function $f$ is denoted by $f(s_1,s_2)$ and ${\bf s}_0=(1,1)$.
Noting that $a$ is a constant, one can integrate Eq. (\ref{partialV}) with respect to $V$ from $V_*$ to $V$ and obtain
\begin{equation}
\label{partial1}
\mu(V,\theta_*)-\mu(V_*,\theta_*)=a\log\frac{V}{V_*}.
\end{equation}
In the same manner, the following equation can be easily derived.
\begin{eqnarray}
\label{partial2}
\mu(V,\theta)-\mu(V,\theta_*)=b\log\frac{\theta}{\theta_*},\\
b\equiv\left.\frac{\partial f(s_1,s_2)}{\partial s_2}\right|_{\bf s_0},
\end{eqnarray}
Adding Eq. (\ref{partial1}) to Eq. (\ref{partial2}) leads to Eq. (\ref{RSF_1state}).
Namely, Eq. (\ref{start1}) is equivalent to Eq. (\ref{RSF_1state}).

\subsubsection{Another example}
Note that Eq. (\ref{start1}) is not the only consequence of the dimensional analysis.
Alternatively, one may choose the ratio of friction coefficients measured at two different states;
i.e., $F(x,y) = x/y$.
Then Eq. (\ref{generalform}) leads to
\begin{equation}
\label{ratio}
\frac{\mu(V,\theta)}{\mu(V_*,\theta_*)} = f\left(\frac{V}{V_*},\frac{\theta}{\theta_*}\right).
\end{equation}
From this equation, one can easily derive 
\begin{equation}
\label{powerlaw}
\mu(V,\theta) = \mu(V_*,\theta_*) 
\left(\frac{V}{V_*}\right)^{\alpha} \left(\frac{\theta}{\theta_*}\right)^{\beta},
\end{equation}
where the exponents $\alpha$ and $\beta$ are constants.

\subsection{Summary of one state-variable case}
A choice of $F(\cdot,\cdot)$ in Eq. (\ref{generalform}) is arbitrary, but it determines the functional form of friction law inevitably.
If we choose to describe the friction coefficient as the difference from the reference state,
the logarithmic form of friction law is inevitable as a direct consequence of dimensional analysis.
This constitutes our central result.
Other choices are also possible, but they yield different forms for the friction law.

\section{The case of two state variables}
The RSF law is sometimes extended to include a second state variable.
Hereafter the second state variable is denoted by $\phi$.
We assume that the friction coefficient is uniquely determined by these three parameters.
Namely, we can write
\begin{equation}
\mu=\mu(V, \theta, \phi).
\end{equation}
The possible form of the friction law depends entirely on the dimension of the second state variable, $\phi$.
(Here we assume that $\theta$ has the dimension of time.)
We discuss two cases separately.

\subsection{If $\phi$ has the dimension of time}
\label{samedimension}
First we discuss the case in which $\phi$ has the dimension of time.
As the dimension of $\theta$ is also time, the ratio $\theta/\phi$ is nondimensional.
Then the $\Pi$ theorem concludes $\mu=f(\theta/\phi)$, where $f$ is an arbitrary function.
In this case, the evolution laws for $\theta$ and $\phi$ must be chosen appropriately
so as to fit experimental data.
However, we are unaware of such a formulation of friction law.

Alternatively, one may again introduce a reference state, which is denoted by $(V_*, \theta_*, \phi_*)$.
In this case, there are actually four independent nondimensional numbers:
 $V/V_*$, $\theta/\theta_*$, $\phi/\phi_*$, $\theta/\phi$.
Then following the same logic as in the previous section, one is led to
\begin{equation}
\label{start2}
\mu(V,\theta,\phi)-\mu(V_*,\theta_*,\phi_*)=f\left(
\frac{V}{V_*},\frac{\theta}{\theta_*},\frac{\phi}{\phi_*},\frac{\theta}{\phi}\right).
\end{equation}
The scaling function $f$ on the right hand side will be denoted by $f({\bf s})$ or $f(s_1, \cdots, s_4)$.
Then, differentiating the both sides of Eq. (\ref{start2}) with respect to $\theta$, and letting $V=V_*$, $\theta=\theta_*$, and $\phi=\phi_*$, 
one obtains
\begin{equation}
\label{wrtTheta}
\frac{\partial \mu(V,\theta,\phi)}{\partial\theta} =
\frac{1}{\theta}\left.\frac{\partial f}{\partial s_2}\right|_{{\bf s}_0}
+ \frac{1}{\phi}\left.\frac{\partial f}{\partial s_4}\right|_{{\bf s}_0},
\end{equation}
where ${\bf s_0}=(1,1,1,\theta/\phi)$.
Note that the coefficients $\partial f/\partial s_2|_{{\bf s}_0}$ and $\partial f/\partial s_4|_{{\bf s}_0}$ depend on $\theta/\phi$.
Therefore, one cannot conclude $\mu(V,\theta,\phi)-\mu(V,\theta_*,\phi)\propto \log(\theta/\theta_*)$ from this equation.

One can repeat the same discussion with respect to $\phi$ and has
\begin{equation}
\label{wrtPhi}
\frac{\partial \mu(V,\theta,\phi)}{\partial\phi} =
\frac{1}{\phi}\left.\frac{\partial f}{\partial s_3}\right|_{{\bf s}_0}
- \frac{\theta}{\phi^2}\left.\frac{\partial f}{\partial s_4}\right|_{{\bf s}_0}.
\end{equation}
Again, this equation does not lead to $\mu(V,\theta,\phi)-\mu(V,\theta,\phi_*)\propto\log(\phi/\phi_*)$.

Contrastingly, the following equation can be obtained with respect to $V$.
\begin{equation}
\label{wrtV}
\frac{\partial\mu(V,\theta,\phi)}{\partial V}=\frac{1}{V}\left.\frac{\partial f}{\partial s_1}\right|_{{\bf s}_0},
\end{equation}
and therefore 
\begin{eqnarray}
\label{wrtV}
\mu(V,\theta,\phi)-\mu(V_*,\theta,\phi)=a\left(\frac{\theta}{\phi}\right)\log\frac{V}{V_*},\\
a\left(\frac{\theta}{\phi}\right) \equiv \left.\frac{\partial f}{\partial s_1}\right|_{{\bf s}_0}.
\end{eqnarray}

Taking Eqs. (\ref{wrtTheta}), (\ref{wrtPhi}), and (\ref{wrtV}) into account, we obtain
\begin{equation}
\label{twostateRSF_true}
\mu(V,\theta,\phi)-\mu(V_*,\theta_*,\phi_*)= a\left(\frac{\theta}{\phi}\right)\log\frac{V}{V_*}
+G\left(\frac{\theta}{\theta_*},\frac{\phi}{\phi_*},\frac{\theta}{\phi}\right),
\end{equation}
where $G$ is an arbitrary function.

Two state variable RSF law is often assumed as the following form \cite{gu1984,blanpied1986,noda2012}.
\begin{equation}
\label{twostateRSF_false}
\mu(V,\theta,\phi)-\mu(V_*,\theta_*,\phi_*)=a\log\frac{V}{V_*}+b_1\log\frac{\theta}{\theta_*}+b_2\log\frac{\phi}{\phi_*}.
\end{equation}
Note that this is just a special case of Eq. (\ref{twostateRSF_true}).
One may thus assume Eq. (\ref{twostateRSF_false}) as long as it can describe experimental data well, because the ultimate justification of friction laws relies entirely on the comparison with experiments.
Here we just point out that Eq. (\ref{twostateRSF_true}) is more general than Eq. (\ref{twostateRSF_false}) from the viewpoint of dimensional analysis.

\subsection{\label{independent_2nd_state} If the dimension of $\phi$ is independent of $V$ and $\theta$}
We assume that there exist no set of integers $(n, m, l)$ that $V^n\theta^m\phi^l$ is nondimensional.
If such integers exist, the $\Pi$ theorem concludes $\mu=f(V^n\theta^m\phi^n)$,
and no restriction is assigned to its functional form from dimensional analysis only.

If such integers do not exist, one may write the analogue of Eq. (\ref{generalform}) from dimensional consistency.
\begin{equation}
\mu(V,\theta,\phi)-\mu(V_*,\theta_*,\phi_*) = f\left(\frac{V}{V_*},\frac{\theta}{\theta_*},\frac{\phi}{\phi_*}\right).
\end{equation}
Following the same discussion as the one state variable case, the above equation leads to Eq. (\ref{twostateRSF_false}).
Therefore, Eq. (\ref{twostateRSF_false}) is a general form of friction law if one cannot make a nondimensional number using $V$, $\theta$, and $\phi$.

An example of such $\phi$ is the temperature, the stresses, or the mass density.
Particularly, adopting the temperature as the second state variable may be plausible in modeling phenomena in which frictional heating is relevant.

\section{Three or more state variables}
The state variables are denoted by $\theta_i$.
Then, if there exist no such integers $(n,m_i)$ that $V^n\prod\theta_i^{m_i}$ is nondimensional, 
and if no intrinsic physical constants are known, 
the deviation of the friction coefficient from a reference state, $\mu(V^*,\{\theta_i^*\})$, must be written as
\begin{equation}
\label{threestates}
\mu(V,\{\theta_i\})-\mu(V^*,\{\theta_i^*\})=a\log\frac{V}{V^*}+\sum b_i\log\frac{\theta_i}{\theta_i^*}.
\end{equation}
However, such choice of state variables are generally impossible if the number of state variables is larger than three, as the number of independent dimensions are typically four: mass, time, length, and temperature.
If there are four state variables, all the variables including the velocity cannot be independent and therefore Eq. (\ref{threestates}) does not necessarily hold.
Then, as in section \ref{samedimension}, no restriction is imposed to the functional form of friction law from dimensional analysis alone.

\section{Conclusions}
In this paper, with the aid of dimensional analysis, it is shown that the logarithmic terms in the RSF law are inevitable irrespective of the actual physical processes if one describes the friction coefficient in terms of the difference from the reference-state value.
With regard to the two state-variable formulation, the logarithmic terms must be interpreted as an additional assumption unless two state variables have independent dimensions.

In general, the formulation of empirical laws with a reference state is rather natural and universal when one cannot know the relevant physical processes.
Therefore, the formulation presented here may be still useful in describing novel frictional properties,
for example, significant weakening behaviors at intermediate to high slip velocities, mm/s to m/s \cite{goldsby2002,han2007,han2011,diToro2011}.
In these phenomena, the second state variable may represent a local temperature at microscopic junctions \cite{rice2006}, the mass density of silica gel \cite{goldsby2002}, or the density of nanopowders \cite{han2011}.
In each case, the dimension of a second state variable is independent of time.
Therefore, the discussion given in section \ref{independent_2nd_state} applies and we may assume Eq. (\ref{twostateRSF_false}) as the empirical friction law.

We conclude that the logarithmic term is derived if (1) one does not know an intrinsic velocity constant and uses a reference state velocity, and if (2) the friction coefficient is expressed as the difference from a reference state.
Then, the contraposition is the following:
A non-logarithmic behaviour implies that (3) an intrinsic velocity constant is used and a reference state velocity is ruled out, or (4) the friction coefficient is not expressed as the difference from a reference-state value.
For instance, a velocity-weakening behaviour in some experiments at high slip velocities \cite{diToro2011,kuwano2011} might be described as 
\begin{equation}
\label{FH}
\mu(V)=\mu_0 - \Delta\mu \frac{V_w}{V},
\end{equation}
where $\mu_0$, $\Delta\mu$, and $V_w$ are constants \cite{rice2006}.
This equation contains the intrinsic velocity constant $V_w$, and therefore corresponds to case (3) above.
An example for case (4) is Eqs.(\ref{ratio}) and (\ref{powerlaw}), where the friction coefficient is expressed as the ratio to a reference-state value.

\begin{acknowledgments}
The author acknowledges Hiroyuki Noda for critical reading of the manuscript.
He also acknowledges JSPS Grand-in-Aid for Scientific Research, {\it New Perspective of the Great Subduction Zone Earthquake from the Super Deep Drilling}.
\end{acknowledgments}




\label{lastpage}
\end{document}